\documentclass[12pt]{iopart}

\usepackage{color}
\usepackage{graphicx}
\usepackage{float}
\usepackage{epsfig}
\usepackage{epstopdf}
\usepackage[caption=false]{subfig}
\usepackage{bm} 
\usepackage{hyperref}

\def\bzeta{{\bm \zeta}}

\def\bzetahat{\widehat \bzeta}

\def\calO{{\cal O}}

\def\del2{\nabla^2}

\def\fsAver#1{\left\langle #1 \right\rangle} 

\def\begeqn{\begin{equation}}
\def\endeqn{\end{equation}}
\def\begeqnar{\begin{eqnarray}}
\def\endeqnar{\end{eqnarray}}
\def\begeqnarn{\begin{eqnarray*}}
\def\endeqnarn{\end{eqnarray*}}

\def\bRhat{\widehat {\bm R}}

\begin{document}
\title{Role of explosive instabilities in high-$\beta$ disruptions in tokamaks}
\author{A. Y. Aydemir, H. H. Lee, S. G. Lee, J. Seol, B. H. Park, and Y. K. In}
\address{National Fusion Research Institute, Daejeon 305-806, Korea}                        
\eads{aydemir@nfri.re.kr}

\begin{abstract} 
Intrinsically explosive growth of a ballooning finger is demonstrated in nonlinear magnetohydrodynamic calculations of high-$\beta$ disruptions in tokamaks. The explosive finger  is formed  by an ideally unstable $n=1$ mode, dominated by an $m/n=2/1$ component. The quadrupole geometry of the $2/1$ perturbed pressure field provides a generic mechanism for  the formation of the initial ballooning finger and its subsequent transition from exponential to explosive growth, without relying on secondary processes.  The explosive ejection of the hot plasma from the core and stochastization of the magnetic field occur in Alfv\'enic time scales, accounting for the extremely fast growth of the precursor oscillations and the rapidity of the thermal quench in  some high-$\beta$ disruptions. 
\end{abstract}

\begin{center}
\end{center}
\maketitle

Tokamak plasmas operate within imperfectly-defined operational boundaries. Accidental or deliberate crossing of these boundaries generally leads to a sudden loss of confinement: the thermal energy content of the plasma is lost on sub-millisecond time scales (thermal quench), followed by a somewhat slower (ms time scale) loss of poloidal magnetic energy (current quench) (see, for example, Wesson\cite{wessonBook2004}, Sect. 7.7, for a basic discussion, and Hender\cite{hender2007} for more recent developments). The energy lost from the plasma invariably ends up on plasma facing components such as the first wall or divertor tiles.  This sudden energy flux, and the forces generated on the surrounding structures by currents that directly couple to them or are inductively induced during the current quench,  pose a significant threat to the structural integrity of the device\cite{aydemir2000, strauss2013}. Hence, understanding disruptions, and their prevention or mitigation continue to be very active research areas in magnetic fusion.

Causes of disruptions in tokamaks are generally divided into four major categories. (i)~Density-limit disruptions are caused by plasma density that exceeds a mostly\break experimentally-defined ``Greenwald limit,'' $n_{GW}$\cite{greenwald2002}. (ii)~Current-limit disruptions  are caused by a macroscopic magnetohydrodynamic (MHD) instability when the plasma current $I_p$ is too high and results in an edge safety factor that is too low,  $q_{e} < 2$\cite{shafranov1970, wesson1978}. Here $q_e=aB_t/RB_p$ (in cylindrical geometry) where $B_p,~B_t$ are the poloidal and toroidal field strengths, respectively, and $a,~R$ are the minor and major radii of the torus. (iii)~$\beta$-limit disruptions result when the plasma pressure is too high, typically represented as a limit on the normalized (dimensional) quantity $\beta_N = \beta(aB_t)/I_p,$ where $\beta=2\mu_0\fsAver{p}/B_t^2$ is the ratio of the plasma pressure to the magnetic pressure, and $\fsAver{..}$ represents a volume average. (iv)~In the last category are vertical displacement events (VDE's) that result from a loss of positional control that brings the plasma in contact with the material structures around it (see, for example, Granetz\cite{granetz1996}). All disruptive events tend to be accompanied by an eventual loss of vertical control; thus, a VDE may be the cause or result of a disruption due to other reasons.

Fusion power density $P_F$ depends quadratically on the electron density $n_e$. Since the magnetic field strength is generally limited by engineering concerns, high fusion power requires high $\beta$ values ($P_F \propto \beta^2$), thus making $\beta$-limit disruptions particularly important for present-day tokamaks and the next step device ITER. 

Some disruptions at high-$\beta$ tend to have very fast-growing precursors or no discernible precursors at all\cite{fredrickson1995, yoshino2005}. When they are present, the precursor oscillations, for example on the edge magnetic field signals,  precede the thermal quench typically by less than a millisecond; time evolution of the signals suggests a mode growing at least exponentially and possibly explosively (faster than exponentially) well into the nonlinear regime.

Hastie, in the context of sawtooth oscillations\cite{hastie1987}, and Callen, for high-$\beta$ disruptions\cite{callen1999}, showed that an ideal MHD mode going through its marginal point as the plasma $\beta$ increases linearly due to external heating may grow as $\exp{(t/\tau)^{3/2}}$, thus providing a plausible explanation for the short time-scale of the precursors. In both works, the characteristic time $\tau \propto (\tau_A^2\tau_h)^{1/3}$ is a geometric mean of the MHD ($\tau_A$) and heating ($\tau_h$) time scales. Under these conditions, faster than exponential growth of an ideal $n=1$ mode was confirmed in numerical simulations of a particular DIII-D disruption (see, for example, \cite{chu1996}) using the NIMROD code\cite{kruger2005}. However, this simulation seems to have used only the $n=0,1$ modes in the toroidal Fourier expansions; thus, it is only a ``quasi-linear,'' not a proper nonlinear calculation, which would require the presence of higher harmonics of the $n=1$ mode. In addition, this particular path to explosive growth was shown to be implausible, requiring an unrealistically small initial perturbation amplitude to generate Alfv\'enic growth, due to the vast separation in time scales $\tau_A$ and $\tau_h$\cite{cowley1997}.

Some characteristics of the actual MHD modes possibly involved in high-$\beta$ disruptions were studied by Park\cite{park1995} and Kleva\cite{kleva1998, kleva1999}. In the former a high pressure front produced by a non-resonant $m/n=1/1$ mode (the quasi-interchange mode\cite{wesson1986, aydemir1987a}) was shown to be unstable to high-$n$ ballooning modes, which nonlinearly lead to a disruption. Kleva's work relied on resistive ballooning modes or their collisionless analogue and was challenged\cite{connor1999} as being not relevant much below the ideal ballooning stability limit (see also \cite{drake1985}).

In this Letter, we show computationally that the ideal $m/n=2/1$ mode, because of its unique geometric characteristics, may provide a generic mechanism for an explosive instability directly and thus may be responsible for  those  high-$\beta$ disruptions with very fast precursors. The numerical tool used is the CTD code (see \cite{aydemir2015} and the references therein) that solves the nonlinear, resistive MHD equations in toroidal geometry. In CTD the poloidal and toroidal directions are treated spectrally while a finite-difference mesh is used in the radial direction. Noncircular geometries are treated through a conformal transformation from the poloidal coordinates $(r,\theta)$  to the unit circle in conformal coordinates $(\rho_c,\theta_c)$ \cite{aydemir1990}. A resistive wall\cite{fitzpatrick1996} and diverted plasmas\cite{aydemir2007b} can also be treated, but for computational efficiency the results discussed here will be in circular geometry with an ideal wall. However, the coordinate axis will be shifted to the magnetic axis with a conformal transformation to approximately align the $\rho_c=const.$ surfaces with flux surfaces. More general results will be presented elsewhere.

Experimental context for these calculations is, ironically, not disruptions but highly stable discharges that last many tens of seconds while exhibiting a long-lived $2/1$ mode in the KSTAR tokamak\cite{lee2015}, under conditions similar to ``advanced/hybrid scenarios''\cite{sips2005},  with high edge safety factor $q_{q5}$ and a low inductive current fraction. During investigations into the nature of this mode, it was discovered that in equilibria with peaked pressure profiles, either a benign $2/1$ mode that saturates at a small amplitude, or an explosively growing $2/1$ is possible, depending on the details of the pressure profile. Leaving a more comprehensive treatment to a future publication, this Letter will focus only on the explosively growing mode.

\begin{figure}[htbp]
\begin{center}
\includegraphics[width=5in]{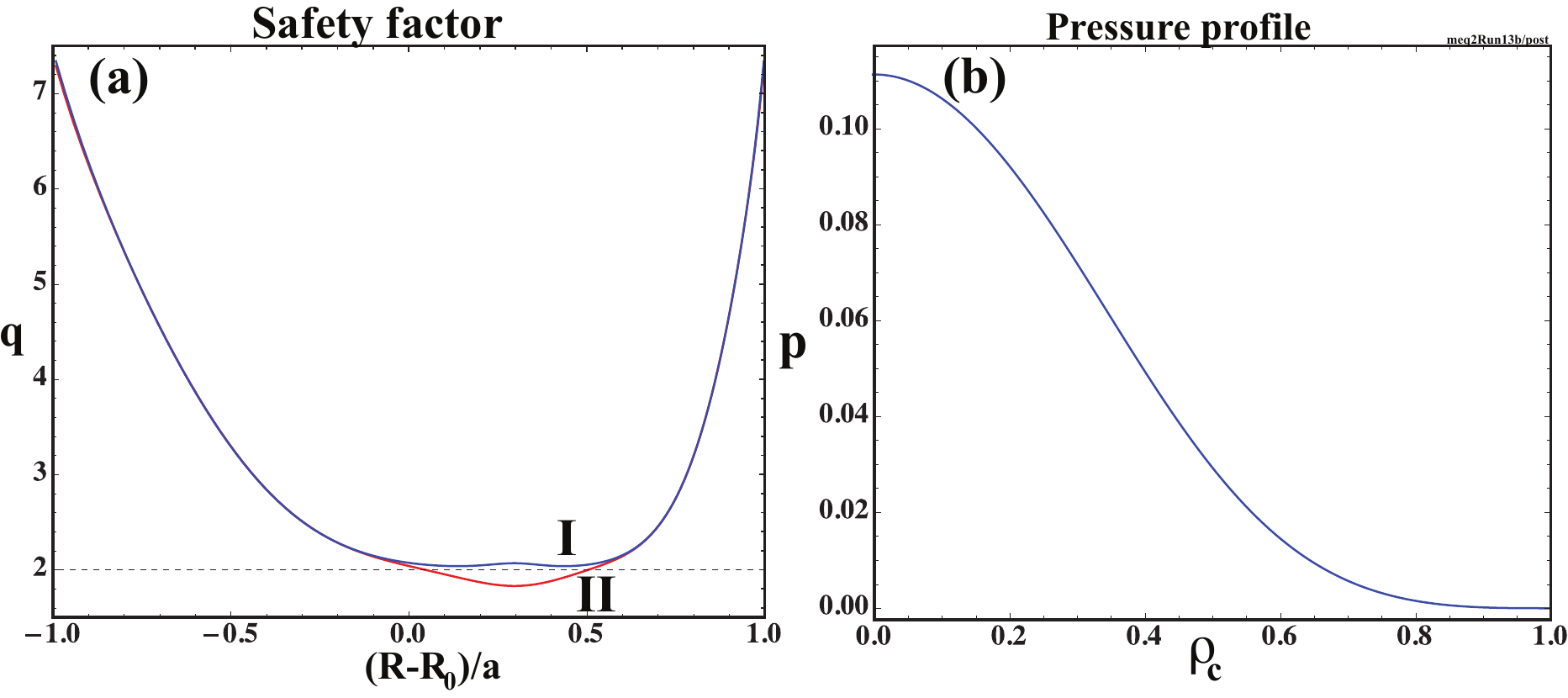}
\caption{\em \baselineskip 14pt Equilibirum profiles. The $q$-profiles in (a) are calculated by field-line tracing and are shown as a function of $(R-R_0)/a$, where $a,R_0$ are the minor and major radius, respectively. The slightly non-monotonic profile $q_I$ has $q_{min}=2.04,~q_0=2.11.$ The monotonic profile $q_{II}$ has $q_0=1.83.$ For both profiles, $q_{95}=7.3.$ The dashed line is at  $q=2.$ (b) The pressure profile at the mid-plane as a function of $\rho_c,$ the normalized radial variable of the conformal coordinate system $(\rho_c,\theta_c).$}
\label{fig:qAndP}
\end{center}
\end{figure}

The equilibrium safety factor and pressure profiles used in the calculations, representative of the advanced/hybrid scenarios in KSTAR,  are shown in Fig.~\ref{fig:qAndP}. The edge and central safety factors are $q_{95} \sim 7, q_0\sim 2$, respectively.  The pressure profiles tend to be peaked due to on-axis electron cyclotron resonance heating (ECRH). Although the plasma-$\beta$ is moderate for the cases considered here ($\beta_N\simeq 1.4$), the $n=1$ mode is ideally unstable for both $q$-profiles shown in Fig.~\ref{fig:qAndP}. The pressure pedestal typical of the H(igh)-mode discharges is avoided numerically to prevent complications introduced by the edge-localized modes (ELM's).

\begin{figure}[htbp]
\begin{center}
\includegraphics[width=5in]{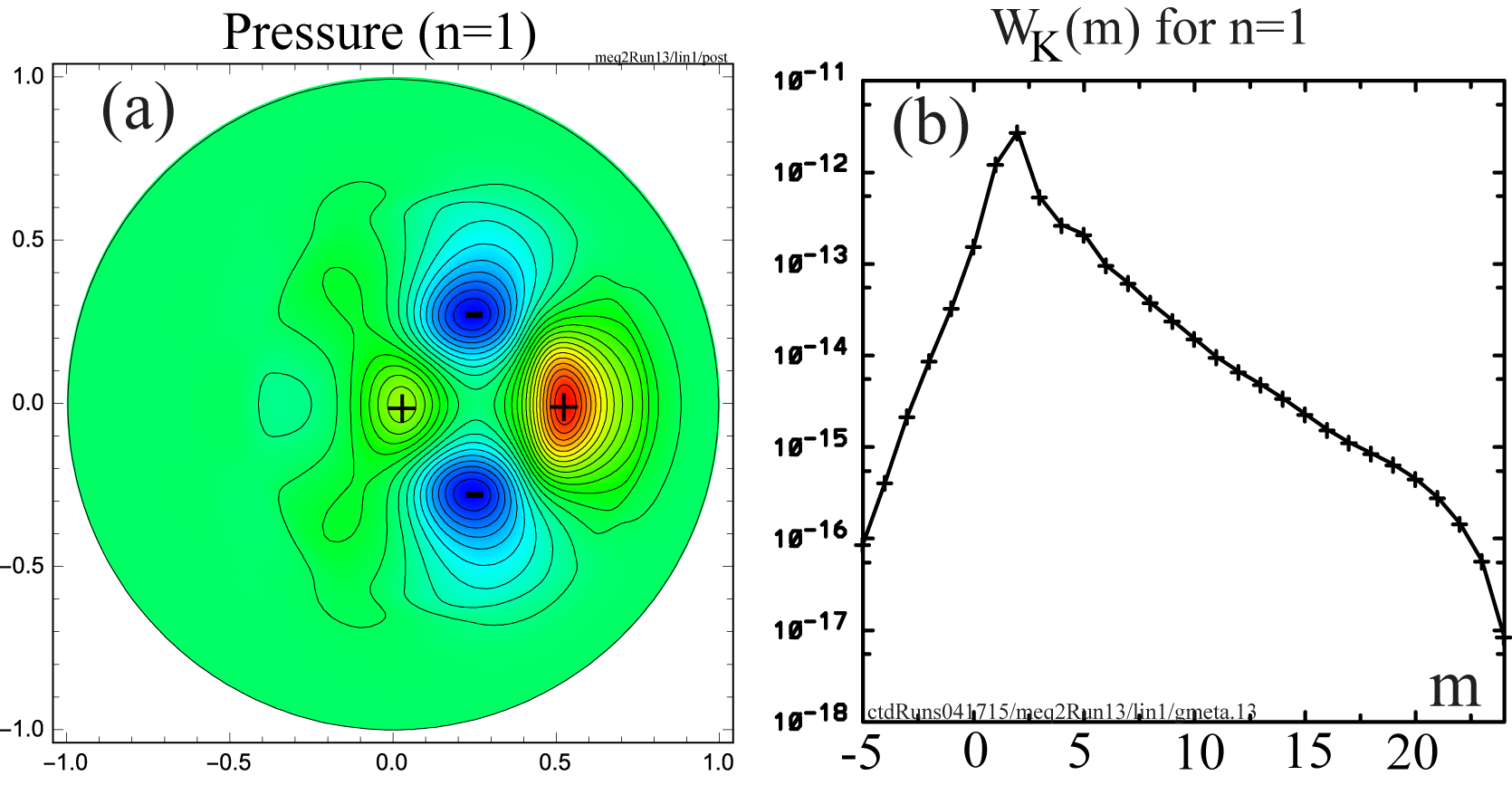}
\caption{\em \baselineskip 14pt (a) Perturbed pressure contours for the $n=1$ mode using the safety factor profile $q_I$ in Fig.~\ref{fig:qAndP} (a). Note the ballooning nature of the eigenfunction. (b) Poloidal spectrum of the kinetic energy for the $n=1$ mode.}
\label{fig:nEq1P}
\end{center}
\end{figure}

The $n=1$ eigenfunction for the pressure field is shown in Fig.~\ref{fig:nEq1P} (a) for the $q_I$ profile ($q_{II}$ gives similar results). As seen in the figure, the poloidal structure of the mode is dominated by an $m=2$ component, although smaller $m=3$ effects are also visible. The high and low-field side asymmetry of the perturbed pressure is indicative of a kink-ballooning mode, a low-$n$ version of infernal modes\cite{manickam1987}.  The ballooning character of the mode is apparent also in its poloidal energy spectrum in Fig.~\ref{fig:nEq1P} (b), showing a coupling that goes much beyond the immediate sidebands at $m=2\pm 1.$

\begin{figure}[htbp]
\begin{center}
\includegraphics[width=6in]{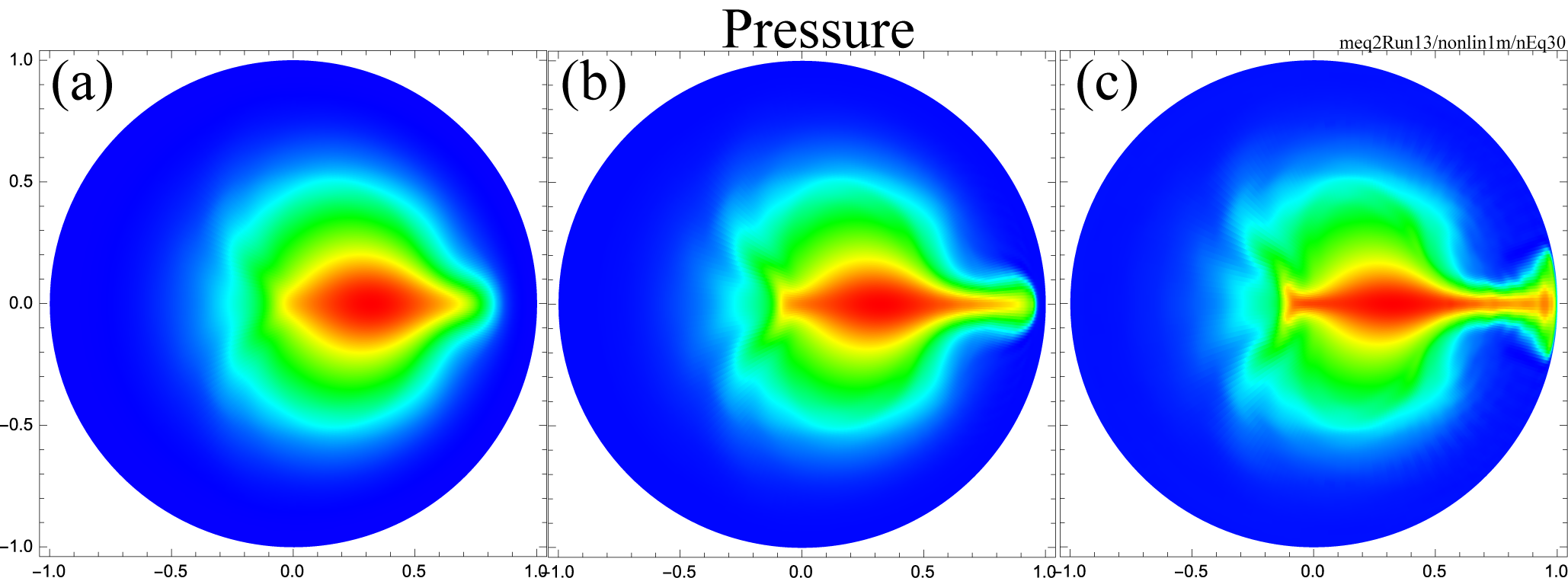}
\caption{\em \baselineskip 14pt Formation and nonlinear evolution of the ``explosive finger'' at the $\zeta=0$ poloidal plane. Figures (a) and (c) are separated by approximately $20$ Alfv\'en times: $t_a = 1483,~t_c=1502$ in units of poloidal Alfv\'en time (See also Fig.~\ref{fig:totalWkHistory}).}
\label{fig:nonlinearP1}
\end{center}
\end{figure}

The quadrupole geometry of the perturbed pressure (due to the dominance of the $m=2$ component) in Fig.~\ref{fig:nEq1P} (a) plays a crucial role in the subsequent nonlinear evolution of the mode. At the toroidal location with this particular phase of the perturbation ($\zeta=0$ poloidal plane in this calculation), the flux and pressure surfaces that are nearly circular initially become elongated in the direction of the major radius ($+\bRhat$) as a result of this perturbation. Continuing elliptical deformation eventually leads to the   formation of a ballooning finger on the low-field side, as seen in Fig.~\ref{fig:nonlinearP1} (a).

Well before the nonlinear evolution of the $n=1$ mode reaches the stage in Fig.~\ref{fig:nonlinearP1} (a), with a well-defined $m=2$ ballooning finger, its growth  becomes explosive. The finger proceeds to get narrower and more elongated, reaching  in $\sim 20$ Alfv\'en times the final stage seen in panel (c), where the hot plasma from the core has splashed against the wall. The end result, of course, is a disruption. 

\begin{figure}[htbp]
\begin{center}
\includegraphics[width=6in]{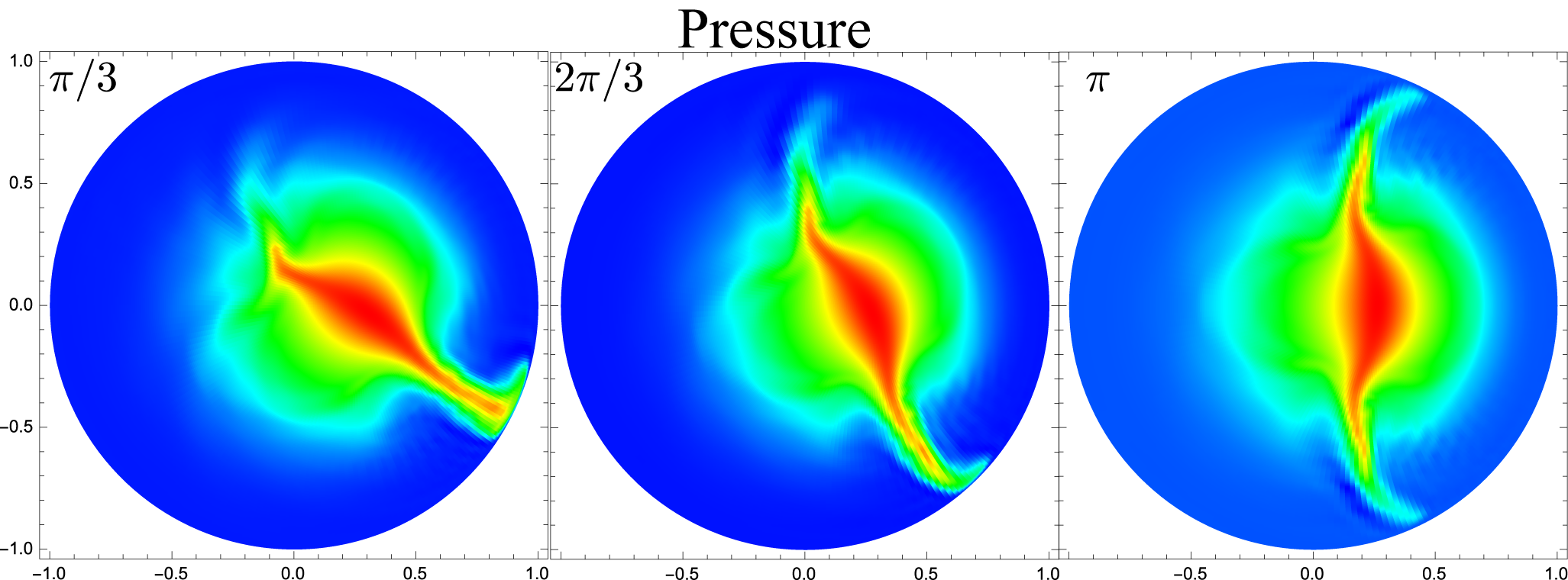}
\caption{\em \baselineskip 14pt Projection of the ballooning finger in Fig.~\ref{fig:nonlinearP1} (c) at $\zeta=0$ to $\zeta=\pi/3, 2\pi/3,$ and $\pi$ poloidal planes. The toroidal field is in the $-\bzetahat$ direction,  resulting in the observed clockwise rotation for increasing toroidal angle $\zeta.$ }
\label{fig:nonlinearP2}
\end{center}
\end{figure}

As stated earlier, the explosively growing ballooning finger and the disruption that accompanies it are driven entirely by a $2/1$ mode; hence, it is not surprising to find an $m=2$ signature in the disruptive phase of the mode.
As seen in Fig.~\ref{fig:nonlinearP2}, projection of the ballooning finger in Fig.~\ref{fig:nonlinearP1}(a) along the field lines to different poloidal planes globally exhibits an approximate 2/1 symmetry. For a linear high-$n$ ballooning mode, the pressure perturbation will satisfy $p \propto f(\zeta-q(r)\theta)g(\theta)h(r)$, where $g(\theta)$ is the ballooning eigenfunction that localizes the mode on the low-field, bad-curvature side of the torus, and h(r) is the radial envelope function. We see that the nonlinear ballooning structure rotates with $q\simeq q_s=2/1$, except near the tip of the finger where the effective $q$ appears to be higher.

\begin{figure}[htbp]
\begin{center}
\includegraphics[width=6in]{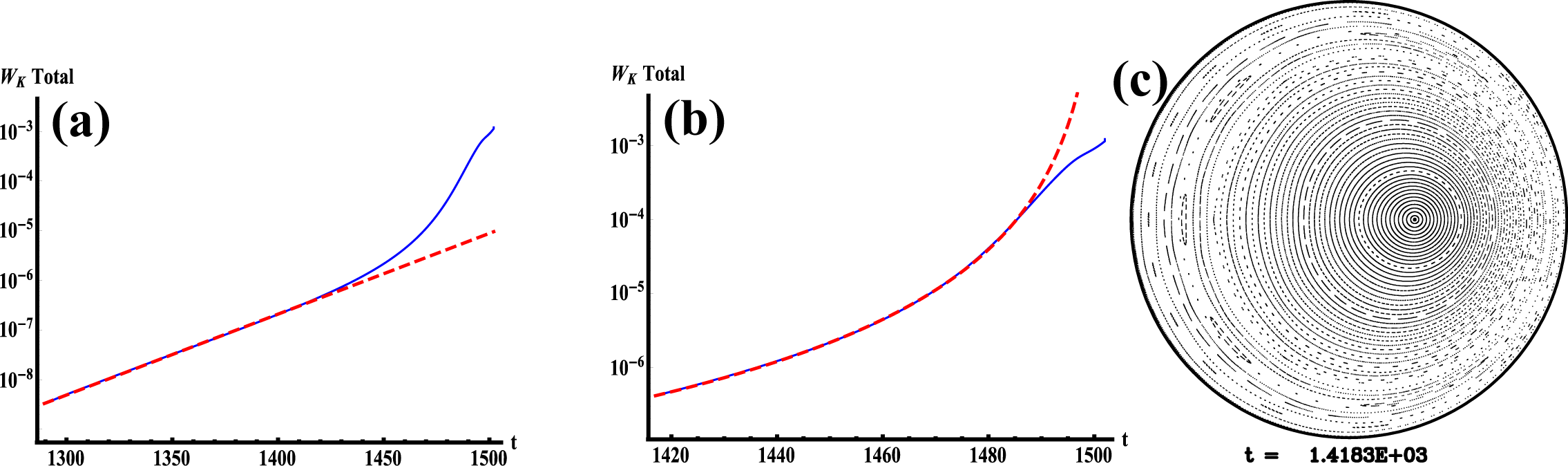}
\caption{\em \baselineskip 14pt (a) The solid curve (blue) shows the growth of the total kinetic energy in $n\ge 1$ modes ($n_{max}=30$ in this calculation). For $t \ge 1420$, the growth rate starts deviating from the purely exponential growth of the dashed line (red). (b) The dashed curve (red) shows a numerical fit to the end phase of the disruption of the form  $W_K(t)=W_K(t_i)[(t_f-t_i)/(t_f-t)]^\nu$ where $\nu=3.37$. (c) Poincar\'e plot of the field lines at the beginning of the explosive growth phase.}
\label{fig:totalWkHistory}
\end{center}
\end{figure}

Explosive nature of the instability is seen in plots of the total kinetic energy (without the $n=0$ contribution) in Fig.~\ref{fig:totalWkHistory}. Panel (a) shows that the  growth rate starts increasing early in the  nonlinear evolution beginning around $t\simeq 1420$ (normalized to the poloidal Alfv\'en time). In this explosive stage, $W_K(t)$ can be fitted with a curve of the form
$W_K(t)=W_K(t_i)[(t_f-t_i)/(t_f-t)]^\nu,$ where $t_i=1417,~t_f=1502,$ and the exponent $\nu=3.37$ (panel (b)) Note that the finite-time singularity model is a very good fit to the numerical results except at the very end when the plasma starts coming in contact with the wall. Panel (c) shows the flux surfaces at the beginning of the nonlinear stage. Except for some high-$m$ resistive modes near the edge, the core plasma is still intact, indicating the small amplitude of the $m=2$ perturbation at this time. This point is confirmed by the energy plots in (a) and (b), which show that at the beginning of the nonlinear behavior, the total kinetic energy is still approximately four orders of magnitude smaller than its final value. Note that the evolution of the explosive finger seen in Fig.~\ref{fig:nonlinearP1} occurs in approximately $20$ Alfv\'en times at the very end of the plots in Fig.~\ref{fig:totalWkHistory}. Thus, this disruption would have only a very brief precursor, much too little to be of practical experimental use.

The discussion up to this point has been entirely about the equilibrium with the safety factor $q_I$ in Fig.~\ref{fig:qAndP}, which has $q>2$ everywhere. The ideal $n=1$ mode displays a very similar behavior with the equilibrium $q$-profile $q_{II}$, which is monotonic and has $q_0=1.83.$ Despite the presence of the $q=2$ rational surface, the $m/n=2/1$ tearing mode is absent; it appears to be stabilized by curvature effects\cite{glasser1975}. Without a significant $m=2$ island, the nonlinear evolution of the $n=1$ mode again proceeds explosively, following a very similar trajectory to the $q_I$ case. This point is illustrated in the brief discussion of the thermal quench below.

\begin{figure}[htbp]
\begin{center}
\includegraphics[width=6in]{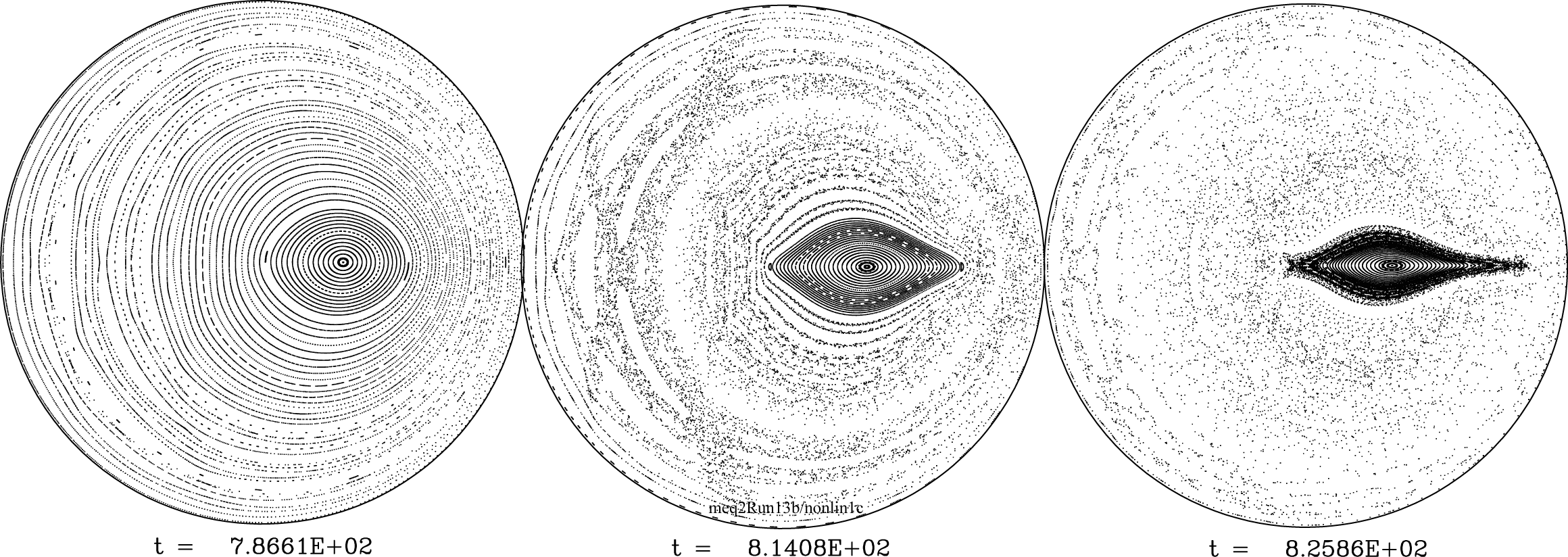}
\caption{\em \baselineskip 14pt Poincar\'e plots of the field lines for a disruption using the equilibrium profile $q_{II}.$ Very similar results are obtained with the $q_I$ profile.}
\label{fig:puncturePlots}
\end{center}
\end{figure}

For both the $q_I$ and $q_{II}$ profiles, the thermal quench has a large convective component in the form of  explosive $m=2$ ballooning fingers. Also on a similar time scale, evolution of the ideal $n=1$ mode  leads  to stochastic field lines, as seen in Fig.~\ref{fig:puncturePlots} from a disruption that started with the $q_{II}$ profile. Thus, parallel thermal transport also plays a significant role during thermal quench, although the loss of energy through this channel will not be explosive but occur on a much slower time scale. Ironically, during the latter stages of the disruption, only the plasma in the explosive finger seems to be well-confined, as the flux surfaces remain intact only inside the finger.

As mentioned earlier, this explosive regime was not the initial goal of our studies, which concentrated on  the ``long-lived'' $2/1$ mode observations in KSTAR\cite{lee2015}. Since it was clear that robustly unstable initial conditions will probably not yield a mode saturated at a small amplitude, as seen in the experiments, our search concentrated on equilibrium profiles near marginal stability where the $n=1$ was only weakly unstable. Thus, the appearance of an eventual explosive regime when the initial linear growth rate  $\gamma\tau_{Ap}\sim \calO(10^{-3})$ or smaller was a surprise. Since then we have observed the same behavior with higher initial growth rates; the high resolution calculations presented here typically started from conditions with  $\gamma\tau_{Ap}\sim \calO(10^{-2}).$

Our numerical results are in qualitative agreement with Cowley's theory\cite{cowley1997, cowley2003} on the explosive growth of ballooning fingers (compare Fig.~2 of Ref.~\cite{cowley2003} with Fig.~\ref{fig:nonlinearP1} (a) above).
But the exponent of the finite-time singularity in the energy $W_K$ in Fig.~\ref{fig:totalWkHistory} (b) ($\nu=3.37$) differs  by nearly a factor of two from $\nu=6.4$ found in Cowley's original work in slab geometry. However, a subsequent analysis in full toroidal geometry by Wilson and Cowley\cite{wilson2004} predicts that the exponent $\nu_d$ for the finite-time singularity in the displacement $\xi$ depends on the Mercier indices $\lambda_{L,S} = 1/2 \mp\sqrt{1/4-D_M},$ with the exponent given by $\nu_d=\lambda_S-\lambda_L.$ Assuming $D_M\simeq 0$ in our circular geometry yields $\nu_d=1$ and $\xi \sim 1/(t_f-t).$ Then using $W_K\sim |\partial\xi/\partial t|^2$ leads to the theoretical prediction $W_K \sim 1/(t_f-t)^4.$ Thus our result in Fig.~\ref{fig:totalWkHistory}(b) with $\nu=3.37$ is much closer to the predictions of the toroidal theory. The small discrepancy may be due finite viscous dissipation and the perfectly conducting boundary conditions used in our simulations, although these have not been investigated in any detail.

While there have been other confirmations of the explosive nonlinear behavior (see, for example, Myers\cite{myers2013} and the references therein), Zhu {\em et al.} have failed to see it in their numerical simulations and have attributed this failure to the ``limited range of validity'' of the explosive regime\cite{zhu2009}. This point might indeed be correct, but we would like to point out that in our studies of the $n=1$ mode, the formation of the thin, explosive finger, and approaching the finite-time singularity in time required us to retain a large number of higher harmonics of the $n=1$ mode; we used up to $n_{max}=30$ in our toroidal Fourier expansions. It is not clear whether Ref.~\cite{zhu2009}, which looked at an $n=15$ ballooning mode, used a correspondingly wide toroidal spectrum in their numerical studies.

In summary, inherently explosive growth of an ideal $n=1$ mode-driven ballooning finger, with a predominantly $m/n=2/1$ component, has been demonstrated in nonlinear toroidal simulations. Transition from exponential to explosive growth is driven nonlinearly by the characteristic quadrupole geometry of the $m=2$ pressure perturbation, which encourages the formation of a ballooning finger.  The finger becomes narrower as it explosively pushes through flux surfaces, in qualitative agreement with published theories. This rapid convective loss of mass and energy from the core to the exterior, in some tens of Alfv\'en times, provides an explanation for fast high-$\beta$ disruptions that occur with little or no precursors. Although the thermal quench starts with the explosive ejection of the hot core, it is also aided, on a slower time scale,  by parallel transport along stochastic fields that are generated during the explosive phase. The aim of this Letter was to bring to the attention of the fusion community in a timely manner, using fast and accurate nonlinear calculations in simple circular geometry, a significant and generic mechanism that may be responsible for high-$\beta$ disruptions in tokamaks with very fast precursors. Exploration of more complex geometries with a divertor is left for a more comprehensive work under preparation.

\section*{Acknowledgments}
This work was supported by MSIP, the Korean Ministry of Science, ICT and Future Planning, through  the KSTAR project.
The authors gratefully acknowledge the help of a referee whose extensive comments and criticism made essential contributions to the manuscript.
\section*{References}

\end{document}